\documentclass[a4paper]{article}

\usepackage{INTERSPEECH2021}
\usepackage{listings}
\usepackage{hyperref}
\usepackage{adjustbox}
\usepackage{amsmath}
\usepackage{amsthm}

\usepackage[labelformat=simple]{subcaption}

\usepackage[font=small,labelfont=bf]{caption}
\usepackage{bbm}
\usepackage{pifont}
\usepackage{enumitem}

\usepackage{xcolor}
\usepackage[linesnumbered,ruled,vlined]{algorithm2e}
\allowdisplaybreaks

\SetCommentSty{mycommfont}

\SetKwInput{KwInput}{Input}                
\SetKwInput{KwOutput}{Output}              
\SetKwInput{KwInit}{Init.}              

\newtheorem{theorem}{Theorem}
\newenvironment{sketch}{%
  \proof}{\endproof}
  
\definecolor{mygreen}{rgb}{0,0.6,0}
\definecolor{mygray}{rgb}{0.5,0.5,0.5}
\definecolor{mymauve}{rgb}{0.58,0,0.82}

\title{Reformulating DOVER-Lap Label Mapping as a Graph Partitioning Problem}
\name{Desh Raj$^1$, Sanjeev Khudanpur$^{1,2}$}
\address{
  $^1$Center for Language and Speech Processing \& $^2$Human Language Technology Center of Excellence \\ 
  The Johns Hopkins University, Baltimore, MD 21218, USA}
\email{draj@cs.jhu.edu, khudanpur@jhu.edu}

\begin{document}

\maketitle
\begin{abstract}
We recently proposed DOVER-Lap, a method for combining overlap-aware speaker diarization system outputs. DOVER-Lap improved upon its predecessor DOVER by using a label mapping method based on globally-informed greedy search. In this paper, we analyze this label mapping in the framework of a maximum orthogonal graph partitioning problem, and present three inferences. First, we show that DOVER-Lap label mapping is exponential in the input size, which poses a challenge when combining a large number of hypotheses. We then revisit the DOVER label mapping algorithm and propose a modification which performs similar to DOVER-Lap while being computationally tractable. We also derive an approximation bound for the algorithm in terms of the maximum number of hypotheses speakers. Finally, we describe a randomized local search algorithm which provides a near-optimal $(1-\epsilon)$-approximate solution to the problem with high probability. We empirically demonstrate the effectiveness of our methods on the AMI meeting corpus. Our code is publicly available: \url{https://github.com/desh2608/dover-lap}.
\end{abstract}

\section{Introduction}

Speaker diarization is the task of segmenting speech into homogeneous speaker-specific regions~\cite{Mir2012SpeakerDA,Tranter2006AnOO}. The traditional approach for speaker diarization involves  a clustering of segment-level speaker embeddings, optionally followed by resegmentation~\cite{GarciaRomero2017SpeakerDU}. In the last few years, supervised methods such as region proposal networks (RPN), end-to-end neural diarization (EEND), and target-speaker voice activity detection (TS-VAD) have been proposed which perform overlapping speaker assignment~\cite{Huang2020SpeakerDW,Fujita2019EndtoEndNS,Medennikov2020TargetSpeakerVA}. We refer the reader to Park et al.~\cite{Park2021ARO} for a review of recent advances in speaker diarization using deep learning.

Since machine learning tasks often benefit from an ensemble of systems (e.g., ROVER is a popular technique for combining ASR system outputs~\cite{Fiscus1997APS}), there has been some interest in developing methods that can combine the outputs from these different diarization systems. DOVER (Diarization Output Voting Error Reduction)~\cite{Stolcke2019DoverAM} was the first method introduced for this task. It recognized that since diarization outputs do not have \textit{absolute} speaker identities, in order to make any \textit{voting mechanism} possible, the outputs must be mapped to a common label space --- we informally refer to this as the \textbf{label mapping} problem. DOVER proposed an algorithm for label mapping using pair-wise linear sum assignment, also known as the Hungarian algorithm~\cite{Kuhn1955TheHM}\footnote{This algorithm is used for mapping hypothesis and reference speakers for \textit{diarization error rate} computation.}.
Xiao et al.~\cite{Xiao2020MicrosoftSD} proposed a simple modification of DOVER capable of handling overlapped speech by using an external threshold for per-speaker voting, but they did not modify the label mapping process in the original method. 

\begin{figure}[t]
    \centering
    \includegraphics[width=0.9\linewidth]{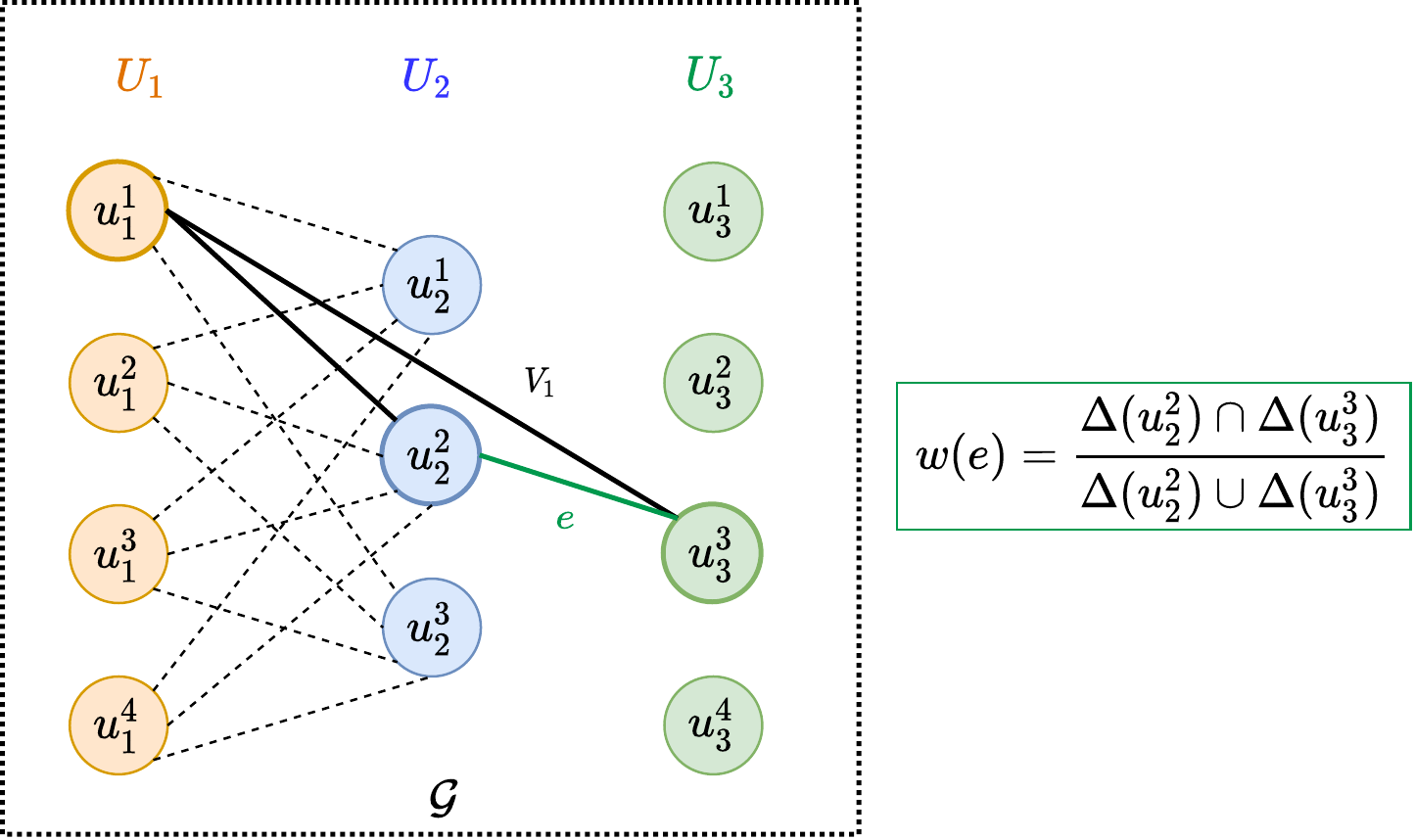}
    \caption{Illustration of the label mapping problem as a graph $\mathcal{G}$ for the case of $K=3$. $V_1$ denotes the clique formed by vertices $(u_1^1,u_2^2,u_3^3)$. $\Delta(u)$ represents the segments where speaker $u$ is active in the recording.}
    \label{fig:label_mapping_graph}
    \vspace{-0.5em}
\end{figure}

In our previous work, we introduced DOVER-Lap, which improved upon DOVER's label mapping by using a globally-informed greedy approximation algorithm~\cite{Raj2020DOVERLapAM}, and empirically demonstrated the strength of this method on the AMI~\cite{Carletta2005TheAM} and LibriCSS~\cite{Chen2020ContinuousSS} datasets, where the combination significantly outperformed the single best system. Due to its ease-of-use and robustness to the underlying hypotheses, DOVER-Lap was successfully used by several top performing systems in the DIHARD-III challenge~\cite{Horiguchi2021TheHD, Wang2021USTCNELSLIPSD}. However, despite the empirical strengths, the algorithms lack analysis or theoretical guarantees, which motivates the questions: \textit{Are there limitations to the performance? Can we design better label mapping algorithms which improve diarization hypothesis combination?}

In this work, we investigate and answer this question by formulating label mapping as a maximum orthogonal partitioning problem. We first show that the DOVER-Lap label mapping algorithm is exponential in the input size, and becomes intractable beyond ensembles of a few systems. We then revisit the pair-wise Hungarian algorithm used in DOVER, and modify it to perform competitively with DOVER-Lap while also being poly-time solvable. We also derive an approximation bound for this algorithm in terms of the maximum number of speakers per hypotheses. Finally, we propose a new label mapping method based on randomized local search which obtains near-optimal performance in theory, and outperforms the deterministic Hungarian method in practice.

\begin{table}[t]
\centering
\caption{Notations used in the paper.}
\label{tab:notation}
\begin{adjustbox}{width=\linewidth} 
\begin{tabular}{@{}cl@{}}
\toprule
\textbf{Symbol} & \textbf{Definition} \\
\midrule
$U_1,\ldots,U_K$ & Diarization hypotheses \\
$c_k$ & Number of speakers in hypotheses $k$ \\
$C$ & Maximum number of speakers in any hypothesis \\
$u_k^i$ & Speaker $i$ in hypothesis $k$ \\
$V$ & Set of all hypotheses speakers \\
$E$ & Set of all edges $\{(u_k^{i},u_{\kappa}^{j})\}$ \\
$w_e$ & Weight on edge $e = \{(u_k^{i},u_{\kappa}^{j})\}$ \\
$(V_1,\ldots,V_C)$ & Clique set (output of label mapping) \\
$\mathcal{G}$ & Graph formed by vertex set $V$ and edge set $E$ with weights $w$ \\
\bottomrule
\end{tabular}
\end{adjustbox}
\end{table}

\vspace{-0.5em}

\section{Problem Formulation}
\label{sec:formalize}

\begin{figure}[t]
    \centering
    \includegraphics[width=0.7\linewidth]{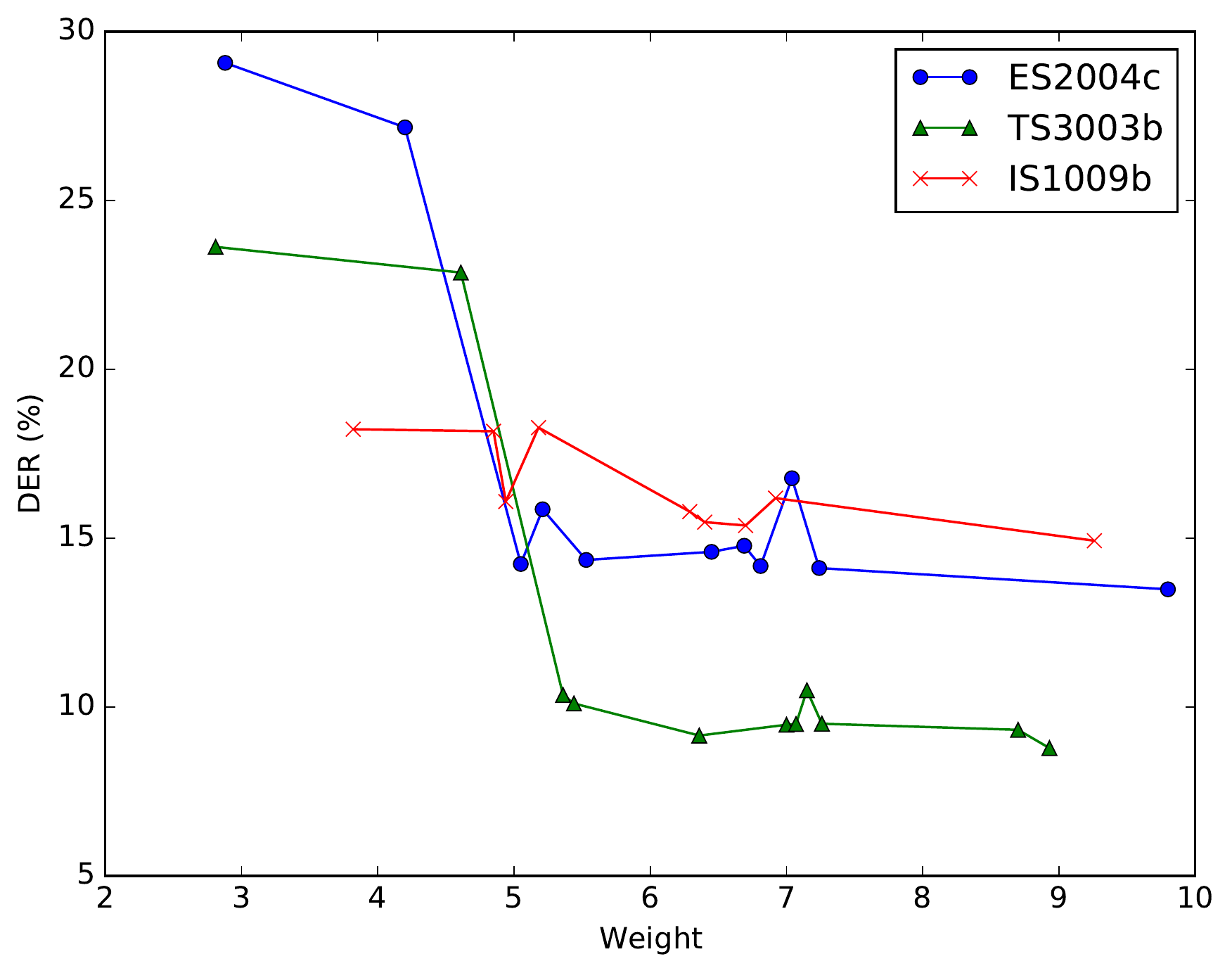}
    \caption{Partition weight $w(\Phi)$ versus diarization error rate (DER) for three arbitrarily chosen recordings from the AMI evaluation set, showing that DER tends to improve with weight.}
    \label{fig:weight_vs_der}
\end{figure}

We will first formalize ``label mapping'' as a graph partitioning problem. For combining diarization outputs, label mapping is the task of relabeling the speakers hypothesized by each component system using a common label set. It is essential for performing region-wise voting among the speakers.

Suppose we have $K$ diarization hypotheses (outputs) $U_1,\ldots,U_K$, containing $c_1,\ldots,c_k$ speakers, respectively, such that $C = \max\{c_k, k\in [K]\}$. Let us denote each speaker as a node, i.e., $u_k^i$ is the node corresponding to the $i^{th}$ speaker in the $k^{th}$ hypothesis, and $V = \{u_k^i\}$ is the set of all speaker nodes. Let $E = \{(u_k^{i},u_{\kappa}^{j}): \forall k,\kappa \in [K], i \in U_k, j\in U_\kappa, k\neq \kappa\}$ denote the set of all edges. Informally, this means that there is an edge between any two nodes if the nodes belong to different hypotheses. Additionally, we have a weight function $w: e \rightarrow \mathbb{R}^+$, where $e$ denotes an edge. In practice, these edge weights are obtained by computing the relative overlap duration between the speakers in the recordings. We define the graph as $\mathcal{G} = (V,E,w)$. It is easy to see that $\mathcal{G}$ is $K$-partite, and if $c_k = C, ~\forall k\in [K]$, then it is also complete. Fig.~\ref{fig:label_mapping_graph} illustrates this graphical formulation of the label mapping problem.

Each $U_k$ in the graph is an \textit{independent set} (set of vertices with no edges between any pair), and the label mapping problem can be defined as: partition $V$ into $C$ vertex-disjoint cliques $\Phi = (V_1,\ldots,V_C)$, such that the partition maximizes
\begin{equation}
\label{eq:objective}
    J(\Phi) = w(\Phi) = \sum_{c\in C} w(V_c) = \sum_{c\in C} \sum_{e\in E(V_c)} w(e),
\end{equation}
where $E(V_c)$ represents edges in the sub-graph induced by $V_c$. Intuitively, the objective maximizes the sum of all edge weights within the cliques.
The partition is \textit{orthogonal} since it may contain at most 1 vertex from every $U_k$. Table~\ref{tab:notation} summarizes these notations.

It may not be immediately clear why maximizing the objective in (\ref{eq:objective}) provides an optimal label mapping. Since the partition is orthogonal, each $V_c$ may represent a mapped speaker label. By maximizing the total edge weights within cliques, we maximize the total relative overlap between speaker turns for speakers that are mapped to the same label. Fig.~\ref{fig:weight_vs_der} illustrates this correspondence between the objective $J$ and the final diarization error rate (DER), validating our conjecture.

\vspace{-0.5em}

\section{A Limitation with DOVER-Lap}

Algorithm~\ref{alg:dl_mapping} presents the label mapping algorithm we proposed earlier~\cite{Raj2020DOVERLapAM}, formalized as the graph problem from Section~\ref{sec:formalize}. The algorithm first computes all the maximal cliques in the graph and greedily chooses from this set (based on edge weight sum) until all the vertices are covered. While this method leads to strong diarization performance by the ensemble, it has a glaring problem --- it is exponential in the number of input hypotheses, since there are $C^K$ maximal cliques in the graph, which we have to compute at line 3 of the algorithm.

\begin{figure}[t]
    \centering
    \includegraphics[width=0.7\linewidth]{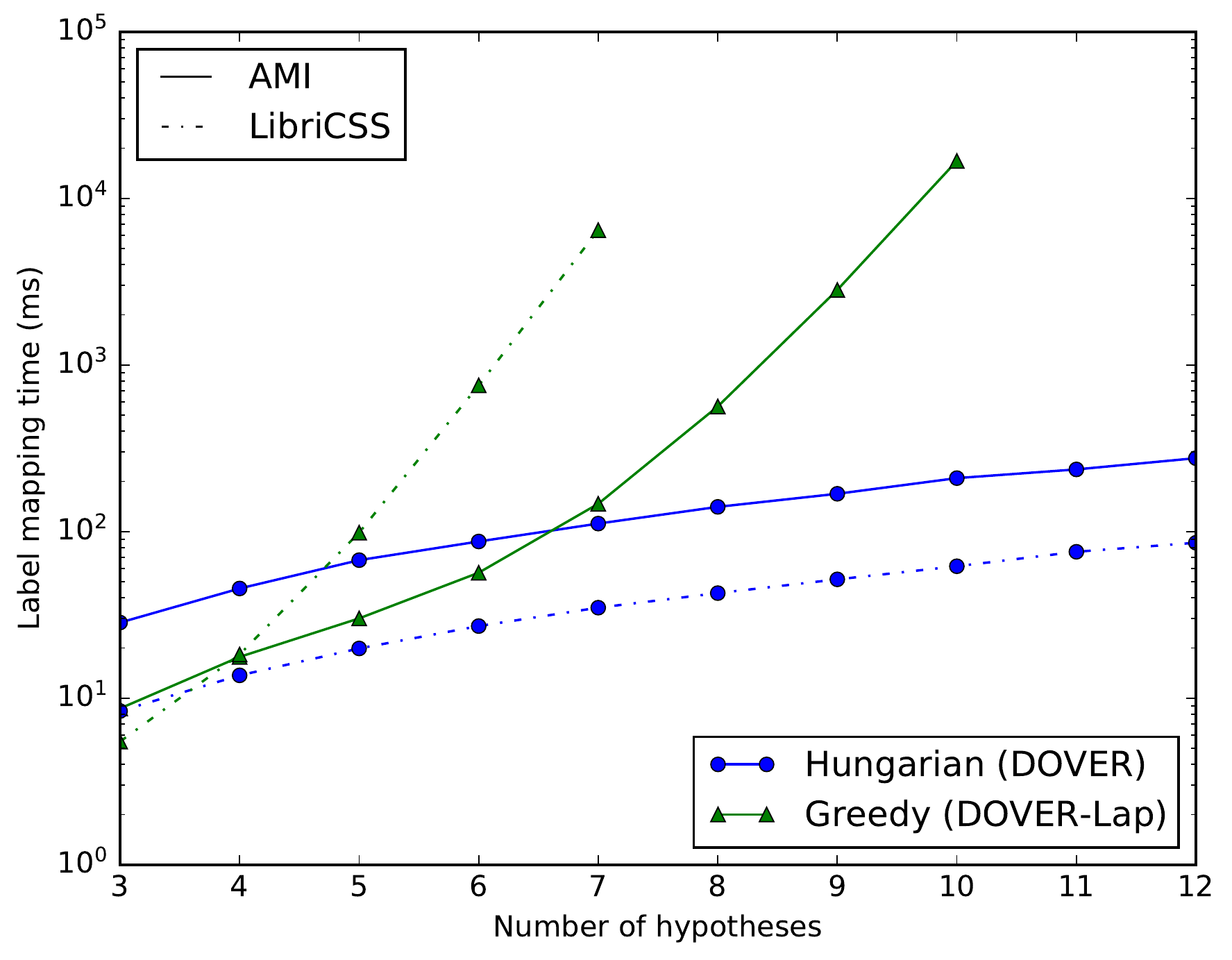}
    \caption{Label mapping time (in ms) for combining different number of hypotheses on the AMI and LibriCSS data. The y-axis is logarithmic.}
    \label{fig:mapping_time}
\end{figure}

Due to this exponential dependency, the algorithm quickly becomes computationally intractable, as shown in Fig.~\ref{fig:mapping_time}. We computed the label mapping time for an increasing number of input hypotheses for the AMI and LibriCSS evaluation sets. For AMI (which contains 4 speakers; solid green line), the algorithm became infeasible beyond $K=10$. For LibriCSS (which contains 8 speakers; dotted green line), this limit was reached for an even smaller value of $K$, making combination impossible beyond 7 hypotheses.

\begin{algorithm}[t]
\DontPrintSemicolon
  
  \KwInput{Graph $\mathcal{G} = (V,E,w)$}
  \KwOutput{Partition $\Phi$ = $V_1,\ldots,V_C$}
  
  $\Phi = \{\}$ 
  
  \tcc{Loop until no vertices remaining}
  \While{$V \neq \phi$}{
    \tcc{Enumerate all maximal cliques}
    $S$ = set of all maximal cliques in $V$
    
    \tcc{Get maximum weighted clique}
    $V_c$ = max($S$, key=$\sum_{e\in S_i} w(e)$) 
    
    \tcc{Add clique to partition}
    $\Phi = \Phi \cup \{V_c\}$
    
    \tcc{Remove clique vertices from V}
    $V = V \setminus \{V_c\}$
  }

\caption{DOVER-Lap label mapping}
\label{alg:dl_mapping}
\end{algorithm}

\vspace{-0.5em}

\section{Revisiting Label Mapping in DOVER}

Although DOVER-Lap improves upon DOVER's label mapping, leading to better DER performance, its exponential complexity creates a bottleneck if we want to combine more diarization systems. In contrast, the DOVER label mapping algorithm is linear in $K$, as evident from Fig.~\ref{fig:mapping_time}. Furthermore, its complexity depends on the number of speaker turns in the hypotheses rather than the number of speakers, which is a desirable property. This trade-off between complexity and performance creates a conundrum: \textit{must we sacrifice diarization performance to obtain tractable computation?} 

To answer this question, we revisit the label mapping algorithm proposed in DOVER, and phrase it in our graph framework. In the next sections, we will describe this redefined formulation, and then show that by making a small modification, we can obtain comparable performance to DOVER-Lap.

\vspace{-0.5em}
\subsection{Base Algorithm}

The DOVER label mapping algorithm is presented in Algorithm~\ref{alg:dover_mapping}. The input hypotheses $U$ are assumed to be in any arbitrary order, but they may also be sorted based on their average diarization ``error'' relative to all other hypotheses. 

The algorithm processes the hypotheses $U$ pair-wise, computing a local map using the Hungarian algorithm, merging the pair, and then updating a global map with the local assignment map. Once all the $U_k$'s are processed, the partitions are obtained by applying the global map on them. The \texttt{merge} operation updates the running hypothesis $\upsilon$ by optionally combining the pair under consideration.

\begin{algorithm}[t]
\DontPrintSemicolon
  
  \KwInput{Graph $\mathcal{G} = (V,E,w)$, $U = \{U_1,\ldots,U_K\}$}
  \KwOutput{Partition $\Phi$ = $V_1,\ldots,V_C$}
  
  $\Psi = \{\}$ \tcc{Global label map}
 
  $\upsilon = U_1$
 
  \For{$k$ in $[2,K]$}{
    \tcc{Compute local map}
    $\psi$ = Hungarian($\upsilon$, $U_k$)
    
    \tcc{Merge pair w.r.t. local map}
    $\upsilon$ = Merge($\upsilon, U_k, \psi$)
    
    \tcc{Update global map}
    $\Psi$ = Update($\Psi,\psi$)
  }
  
  \tcc{Compute partition using global map}
  $\Phi$ = Partition($U$,$\Psi$)
  
\caption{DOVER label mapping}
\label{alg:dover_mapping}
\end{algorithm}

\vspace{-0.5em}
\subsection{Modifying the \texttt{merge} Operation}
\label{sec:dover_implement}

In the original DOVER, the \texttt{merge} operation simply returns the first hypothesis (or the ``anchor''), meaning that $\upsilon$ is fixed throughout the algorithm. An implication of this design is that the starting anchor is crucial for good performance. Stolcke and Yoshioka~\cite{Stolcke2019DoverAM} suggested a workaround for this problem by choosing the anchor based on the average DER to all other hypotheses (i.e., a  hypothesis is scored against all others, and the corresponding DERs are averaged), but even with this DER-based sorting, the ``fixed anchor'' design may be suboptimal. In Table~\ref{tab:dover}, we see that while the sorting improves DER from 30.43\% to 27.95\%, it is still worse than all the systems that are being combined.

We modified the \texttt{merge} operation by allowing it to combine nodes in the two hypotheses if they map to the same speaker, and re-estimating edge weights connecting the merged vertex set to other vertices in $\mathcal{G}$. This edge weight re-estimation can be done in a general graph $\mathcal{G}$ by adding the weights for edges incident on the same mapped vertex. However, for our case, this would over-estimate the weight, so we instead merged the underlying speaker turns in the hypotheses pair to obtain a new hypothesis, which is then used to (re)compute\footnote{In practice, we evaluate edge weights ``on the fly'' when processing hypotheses pairs.} the edge weights to the remaining hypotheses. After this merge operation, the number of independent sets in $\mathcal{G}$ decreases by 1. A pseudocode for the merge operation is as follows:

\begin{lstlisting}[language=python]
def merge(hyp1, hyp2, mapping):
  hyp1_new = map(hyp1, mapping)
  hyp2_new = map(hyp2, mapping)
  merged_hyp = combine_segments(hyp1_new, 
    hyp2_new)
  return merged_hyp
\end{lstlisting}
With this modification in the \texttt{merge} operation, we found that DOVER label mapping obtained competitive performance with DOVER-Lap, as shown in Table~\ref{tab:dover}. We combined the outputs of 3 clustering-based methods: agglomerative hierarchical clustering (AHC)~\cite{GarciaRomero2017SpeakerDU}, spectral clustering (SC)~\cite{Park2020AutoTuningSC}, and Variational Bayes (VBx)~\cite{Dez2019BayesianHB}, where the modified DOVER algorithm outperformed the single-best system. More importantly, sorting the hypotheses in advance was now found to be less critical towards the overall performance of the algorithm, and our modified DOVER (without sorting) outperformed the original DOVER (with sorting) by 5.4\% relative DER. Note that we selected non-overlap-aware hypotheses for this experiment since the original DOVER method assumes single speaker segments.

\begin{table}[t]
\centering
\caption{Comparison of original DOVER implementation and new DOVER with \texttt{merge} operation. Results are shown on the AMI development set, in terms of speaker error (SE) and diarization error rate (DER). We combined 3 hypotheses: agglomerative hierarchical clustering (AHC), spectral clustering (SC), and Variational Bayes (VBx). We used oracle speech segment boundaries, so there are no false alarms. $^\dag$DOVER-Lap has exponential complexity.}
\label{tab:dover}
\begin{adjustbox}{width=0.6\linewidth} 
\begin{tabular}{@{}lcc@{}}
\toprule
\textbf{Method} & \textbf{SE} & \textbf{DER} \\
\midrule
AHC~\cite{GarciaRomero2017SpeakerDU} & 8.29 & 27.82 \\
SC~\cite{Park2020AutoTuningSC} & 6.89 & 26.42 \\
VBx~\cite{Dez2019BayesianHB} & 7.35 & 26.88 \\
\midrule
DOVER (original) & 10.90 & 30.43 \\
 + DER-based sorting & 8.42 & 27.95 \\
\midrule
DOVER (modified) & 6.93 & 26.46 \\
 + DER-based sorting & \textbf{6.77} & \textbf{26.30} \\
DOVER-Lap$^\dag$ & \underline{6.17} & \underline{25.70} \\
\bottomrule
\end{tabular}
\end{adjustbox}
\end{table}

\vspace{-0.5em}
\subsection{Derivation of Approximation Ratio}

We will now derive an approximation bound for our modified DOVER algorithm in terms of the maximum number of speakers $C$. The theorem and proof are based on a similar result presented in He et al.~\cite{He2000ApproximationAF}.

\begin{theorem}
\label{thm:dover}
Algorithm~\ref{alg:dover_mapping} is a $\left(1-\frac{1}{C}\right)$-approximation.
\end{theorem}
\begin{sketch}
We can convert $\mathcal{G}$ into a complete $K$-partite graph without changing the solution, by adding dummy nodes to each $U_k$ until $|U_k|=C, ~\forall k\in [K]$, and 0-weighted edges to all newly added nodes. So W.L.O.G, suppose $\mathcal{G}$ is complete. 

First we will show, by induction on $K$, that $w(\Phi) \geq \frac{w(\mathcal{G})}{C}$. For $K=2$, $\mathcal{G}$ is bipartite, so the Hungarian method provides an optimal solution, and the statement holds by a simple averaging argument. For the inductive case, suppose the statement holds for some $K-1$. Let $\psi_1$ be the matching in the first iteration (i.e., between $U_1$ and $U_2$), and $\Phi^{\prime}$ be the remaining matching. Let $\mathcal{G}^{\prime}$ be the graph obtained after the first merge operation. Then, by applying the statement on $\psi_1$ and $\Phi^{\prime}$, we have
\begin{align*}
w(\Phi) &= w(\psi_1) + w(\Phi^{\prime}) 
        \geq \frac{1}{C} \sum_{e\in [U_1,U_2]}w(e) + \frac{w(\mathcal{G}^{\prime})}{C} \\
        &= \frac{1}{C} \left( \sum_{e\in [U_1,U_2]}w(e) + \sum_{e\notin [U_1,U_2]}w(e) \right) 
        = \frac{w(\mathcal{G})}{C}.
\end{align*}

Now suppose $\Phi^*$ is an optimal solution. Then, since $w(\Phi^*)\leq w(\mathcal{G})$, we have $\frac{w(\Phi)}{w(\Phi^*)} \geq \frac{w(\Phi)}{w(\mathcal{G})} \geq \frac{1}{C}$ (using above). Hence, $1 - \frac{w(\Phi)}{w(\Phi^*)} \leq 1 - \frac{1}{C}$, and so $\left| \frac{w(\Phi^*)-w(\Phi)}{w(\Phi^*)}\right| \leq 1 - \frac{1}{C}$, which proves the theorem.
\end{sketch}

It can be shown (through a reduction from the $c$-way $k$-coloring problem) that there is no efficient (deterministic) algorithm with a better approximation ratio~\cite{He2000ApproximationAF}. This motivates a randomization-based approach, which we will describe next.




\section{Randomized Local Search}

While the DOVER label mapping is fast and effective, it may not provide a good solution when $C$ is large (as indicated by Theorem~\ref{thm:dover}). In this section, we propose a new algorithm for label mapping based on randomized local search (RLS). The entire algorithm is shown in Algorithm~\ref{alg:randomized}.

Each \textit{epoch} of the algorithm initializes $\Phi$ arbitrarily, and then keeps improving the partition using local improvements for a specific number of \textit{iterations} ($M$). The local improvement consists of identifying a high-weighted edge going across cliques in the partition (i.e., in the edge set $E(\Tilde{\Phi}^C)$), and updating the partitions to include it within the clique by swapping either of its incident vertices with another vertex in the same independent set. This procedure is repeated for a fixed number of epochs ($N$), and the partition with the largest weight is returned as output. We state the following theorem, and refer the interested reader to Liu et al.~\cite{Liu2006GeneralizedkmultiwayCP} for a proof.

\begin{algorithm}[t]
\DontPrintSemicolon
  
  \KwInput{Graph $\mathcal{G} = (V,E,w)$}
  \KwOutput{Partition $\Phi$ = $V_1,\ldots,V_C$}
  
  $\Phi = \{\}$
  
  \tcc{Repeat for $N$ epochs}
  \For{$n$ in $[N]$}{
    \tcc{Initialize a partition at random}
    $\Tilde{\Phi}$ = Random($V$)
    
    \tcc{Repeat for $M$ iterations}
    \For{$m$ in $[M]$}{
        
        \tcc{Select edge between cliques}
        $u_p v_q$ = Sample($E(\Tilde{\Phi}^C)$, $p = \frac{w(u_p v_q)}{w(\Tilde{\Phi}^C)}$)
        
        \tcc{Swap incident vertex}
        Swap $u_p$ and $u^{\prime}_p$ with probability $\frac{1}{2}$
        
        Swap $v_q$ and $v^{\prime}_q$ with probability $\frac{1}{2}$
    }
    
    \tcc{Update if weight increases}
    $\Phi$ = max($\Phi$, $\Tilde{\Phi}$)
  }

\caption{Randomized local search}
\label{alg:randomized}
\end{algorithm}

\begin{theorem}
\label{thm:randomized}
For $M=\mathcal{O}(CK)$ and $N$ large enough, Algorithm~\ref{alg:randomized} returns a  $(1-\epsilon)$-approximate solution with probability $\left(1-\frac{1}{e}\right)$.
\end{theorem}


Table~\ref{tab:results} shows the results for the proposed RLS algorithm compared with the modified DOVER label mapping, on the AMI evaluation set. We combined 3 overlap-aware diarization systems: overlap-aware spectral clustering~\cite{Raj2020MulticlassSC}, VB-based overlap assignment~\cite{Bullock2020OverlapawareDR}, and region proposal networks~\cite{Huang2020SpeakerDW}. For RLS, we adopted an early stopping method where we stopped the procedure when the objective function value did not increase for 100 epochs. We used \texttt{spyder}\footnote{\url{https://github.com/desh2608/spyder}} for DER-based sorting in DOVER, and also for evaluating the final performances.

\begin{table}[t]
\centering
\caption{Comparison of the modified DOVER and randomized local search label mapping methods. Results are shown on the AMI evaluation set, in terms of MS, FA, and DER. We combined 3 overlap-aware hypotheses: overlap-aware SC, VB-based overlap assignment, and regional proposal networks (RPN). $^\dag$DOVER-Lap has exponential complexity.}
\label{tab:results}
\begin{adjustbox}{width=\linewidth} 
\begin{tabular}{@{}lcccc@{}}
\toprule
\textbf{Method} & \textbf{MS} & \textbf{FA} & \textbf{SE} & \textbf{DER} \\
\midrule
Overlap-aware SC~\cite{Raj2020MulticlassSC} & 11.48 & 2.27 & 9.81 & 23.56 \\
VB-based overlap assignment~\cite{Bullock2020OverlapawareDR} & 9.84 & \textbf{2.06} & 9.60 & 21.50 \\
Region Proposal Networks~\cite{Huang2020SpeakerDW} & \textbf{9.49} & 7.68 & 8.25 & 25.42 \\
\midrule
DOVER (Modified) & 9.91 & 2.71 & 8.56 & 21.58 \\
 + DER-based sorting & 9.79 & 2.93 & 8.20 & 20.92 \\
Randomized local search (RLS) & 9.69 & 3.21 & \textbf{7.84} & \textbf{20.74} \\
DOVER-Lap$^\dag$ & 9.71 & 3.02 & \underline{7.68} & \underline{20.41} \\
\bottomrule
\end{tabular}
\end{adjustbox}
\end{table}

As expected, RLS outperforms the DOVER algorithm, and most of the gains come from lower speaker error (7.84\% compared with 8.20\% for DOVER). However, this difference in performance is fairly small, especially when we consider that the RLS method requires a longer processing time. This may be because the theoretical bounds are designed to hold in the setting when the size of inputs is fairly large. In our setting of combining diarization hypothesis, these ``large number'' assumptions are violated. Furthermore, improving the objective in (\ref{eq:objective}) is not monotonically related to an improvement in DER, as seen in Fig.~\ref{fig:weight_vs_der}.

Still, these results have important implications. Our experiments with the RLS method indicates that even with a theoretically stronger algorithm, it may not be possible to do much better than the fast Hungarian-based DOVER algorithm (with its modified \texttt{merge} operation), due to the constraints of our setting. As such, it is unlikely that any further advances in combination performances under this framework would be obtained from better label mapping methods. 


\section{Conclusion}

By formulating label mapping as a graph partitioning problem, we showed that the greedy algorithm used in DOVER-Lap becomes intractable as the number of hypotheses increases. We then proposed a modification to the DOVER pair-wise Hungarian method which allows it to empirically perform close to DOVER-Lap while being poly-time solvable. We also derived approximation bounds for this algorithm and showed that it depends on the number of hypotheses speakers. Finally, we proposed a randomized local search method that theoretically and empirically outperformed the deterministic algorithms, although by a small margin. With this analysis in place, we conjectured that no further improvement in system combination may be obtained through better label mapping.

\section{Acknowledgments}
We thank Michael Dinitz for helpful discussions about the approximation ratios, and Paola Garcia, Zili Huang, and Maokui He for providing some of the diarization outputs used in the experiments. This work was partially supported by grants from the JHU Applied Physics Laboratory via the ACHLT2 Program, and the Government of Israel via Project Babylon.

\bibliographystyle{IEEEtran}
\bibliography{main}

\begin{thebibliography}{10}
\providecommand{\url}[1]{#1}
\csname url@samestyle\endcsname
\providecommand{\newblock}{\relax}
\providecommand{\bibinfo}[2]{#2}
\providecommand{\BIBentrySTDinterwordspacing}{\spaceskip=0pt\relax}
\providecommand{\BIBentryALTinterwordstretchfactor}{4}
\providecommand{\BIBentryALTinterwordspacing}{\spaceskip=\fontdimen2\font plus
\BIBentryALTinterwordstretchfactor\fontdimen3\font minus
  \fontdimen4\font\relax}
\providecommand{\BIBforeignlanguage}[2]{{%
\expandafter\ifx\csname l@#1\endcsname\relax
\typeout{** WARNING: IEEEtran.bst: No hyphenation pattern has been}%
\typeout{** loaded for the language `#1'. Using the pattern for}%
\typeout{** the default language instead.}%
\else
\language=\csname l@#1\endcsname
\fi
#2}}
\providecommand{\BIBdecl}{\relax}
\BIBdecl

\bibitem{Mir2012SpeakerDA}
X.~A. Mir{\'o}, S.~Bozonnet, N.~W.~D. Evans, C.~Fredouille, G.~Friedland, and
  O.~Vinyals, ``Speaker diarization: A review of recent research,'' \emph{IEEE
  Transactions on Audio, Speech, and Language Processing}, vol.~20, pp.
  356--370, 2012.

\bibitem{Tranter2006AnOO}
S.~Tranter and D.~A. Reynolds, ``An overview of automatic speaker diarization
  systems,'' \emph{IEEE Transactions on Audio, Speech, and Language
  Processing}, vol.~14, pp. 1557--1565, 2006.

\bibitem{GarciaRomero2017SpeakerDU}
D.~Garcia-Romero, D.~Snyder, G.~Sell, D.~Povey, and A.~McCree, ``Speaker
  diarization using deep neural network embeddings,'' \emph{IEEE International
  Conference on Acoustics, Speech and Signal Processing (ICASSP)}, pp.
  4930--4934, 2017.

\bibitem{Huang2020SpeakerDW}
Z.~Huang, S.~Watanabe, Y.~Fujita, P.~Garc{\'i}a, Y.~Shao, D.~Povey, and
  S.~Khudanpur, ``Speaker diarization with region proposal network,''
  \emph{IEEE International Conference on Acoustics, Speech and Signal
  Processing (ICASSP)}, pp. 6514--6518, 2020.

\bibitem{Fujita2019EndtoEndNS}
Y.~Fujita, N.~Kanda, S.~Horiguchi, Y.~Xue, K.~Nagamatsu, and S.~Watanabe,
  ``End-to-end neural speaker diarization with self-attention,'' \emph{2019
  IEEE Automatic Speech Recognition and Understanding Workshop (ASRU)}, pp.
  296--303, 2019.

\bibitem{Medennikov2020TargetSpeakerVA}
I.~Medennikov, M.~Korenevsky, T.~Prisyach, Y.~Y. Khokhlov, M.~Korenevskaya,
  I.~Sorokin, T.~Timofeeva, A.~Mitrofanov, A.~Andrusenko, I.~Podluzhny,
  A.~Laptev, and A.~Romanenko, ``Target-speaker voice activity detection: a
  novel approach for multi-speaker diarization in a dinner party scenario,''
  \emph{INTERSPEECH}, 2020.

\bibitem{Park2021ARO}
T.~Park, N.~Kanda, D.~Dimitriadis, K.~J. Han, S.~Watanabe, and S.~S. Narayanan,
  ``A review of speaker diarization: Recent advances with deep learning,''
  \emph{ArXiv}, vol. abs/2101.09624, 2021.

\bibitem{Fiscus1997APS}
J.~Fiscus, ``A post-processing system to yield reduced word error rates:
  Recognizer output voting error reduction {(ROVER)},'' \emph{IEEE Workshop on
  Automatic Speech Recognition and Understanding (ASRU)}, pp. 347--354, 1997.

\bibitem{Stolcke2019DoverAM}
A.~Stolcke and T.~Yoshioka, ``{DOVER}: A method for combining diarization
  outputs,'' \emph{2019 IEEE Automatic Speech Recognition and Understanding
  Workshop (ASRU)}, pp. 757--763, 2019.

\bibitem{Kuhn1955TheHM}
H.~W. Kuhn, ``The {Hungarian} method for the assignment problem,'' \emph{Naval
  Research Logistics Quarterly}, vol.~2, pp. 83--97, 1955.

\bibitem{Xiao2020MicrosoftSD}
X.~Xiao, N.~Kanda, Z.~Chen, T.~Zhou, T.~Yoshioka, S.~Chen, Y.~Zhao, G.~Liu,
  Y.~Wu, J.~Wu, S.~Liu, J.~Li, and Y.~Gong, ``Microsoft speaker diarization
  system for the {VoxCeleb} speaker recognition challenge 2020,'' \emph{IEEE
  International Conference on Acoustics, Speech and Signal Processing
  (ICASSP)}, 2021.

\bibitem{Raj2020DOVERLapAM}
D.~Raj, L.~P. Garc{\'i}a-Perera, Z.~Huang, S.~Watanabe, D.~Povey, A.~Stolcke,
  and S.~Khudanpur, ``{DOVER-Lap}: A method for combining overlap-aware
  diarization outputs,'' \emph{IEEE Spoken Language Technology Workshop (SLT)},
  2021.

\bibitem{Carletta2005TheAM}
J.~Carletta, S.~Ashby, S.~Bourban, M.~Flynn, M.~Guillemot, T.~Hain, J.~Kadlec,
  V.~Karaiskos, W.~Kraaij, M.~Kronenthal, G.~Lathoud, M.~Lincoln, A.~L. Masson,
  I.~McCowan, W.~Post, D.~Reidsma, and P.~Wellner, ``The {AMI} meeting corpus:
  A pre-announcement,'' in \emph{MLMI}, 2005.

\bibitem{Chen2020ContinuousSS}
Z.~Chen, T.~Yoshioka, L.~Lu, T.~Zhou, Z.~Meng, Y.~Luo, J.~Wu, and J.~Li,
  ``Continuous speech separation: Dataset and analysis,'' \emph{IEEE
  International Conference on Acoustics, Speech and Signal Processing
  (ICASSP)}, pp. 7284--7288, 2020.

\bibitem{Horiguchi2021TheHD}
S.~Horiguchi, N.~Yalta, P.~Garc{\'i}a, Y.~Takashima, Y.~Xue, D.~Raj, Z.~Huang,
  Y.~Fujita, S.~Watanabe, and S.~Khudanpur, ``The {Hitachi-JHU DIHARD III}
  system: Competitive end-to-end neural diarization and x-vector clustering
  systems combined by {DOVER-Lap},'' \emph{ArXiv}, vol. abs/2102.01363, 2021.

\bibitem{Wang2021USTCNELSLIPSD}
Y.~Wang, M.~He, S.~Niu, L.~Sun, T.~Gao, X.~Fang, J.~Pan, J.~Du, and C.-H. Lee,
  ``{USTC-NELSLIP} system description for {DIHARD-III} challenge,''
  \emph{ArXiv}, vol. abs/2103.10661, 2021.

\bibitem{Park2020AutoTuningSC}
T.~J. Park, K.~J. Han, M.~Kumar, and S.~S. Narayanan, ``Auto-tuning spectral
  clustering for speaker diarization using normalized maximum eigengap,''
  \emph{IEEE Signal Processing Letters}, vol.~27, pp. 381--385, 2020.

\bibitem{Dez2019BayesianHB}
M.~D{\'i}ez, L.~Burget, S.~Wang, J.~Rohdin, and J.~Cernock{\'y}, ``Bayesian hmm
  based x-vector clustering for speaker diarization,'' in \emph{INTERSPEECH},
  2019.

\bibitem{He2000ApproximationAF}
G.~He, J.~Liu, and C.~Zhao, ``Approximation algorithms for some graph
  partitioning problems,'' \emph{J. Graph Algorithms Appl.}, vol.~4, pp. 1--11,
  2000.

\bibitem{Liu2006GeneralizedkmultiwayCP}
J.~Liu, Y.~Peng, and C.~Zhao, ``Generalized k-multiway cut problems,''
  \emph{Journal of Applied Mathematics and Computing}, vol.~21, pp. 69--82,
  2006.

\bibitem{Raj2020MulticlassSC}
D.~Raj, Z.~Huang, and S.~Khudanpur, ``Multi-class spectral clustering with
  overlaps for speaker diarization,'' \emph{IEEE Spoken Language Technology
  Workshop (SLT)}, 2021.

\bibitem{Bullock2020OverlapawareDR}
L.~Bullock, H.~Bredin, and L.~P. Garc{\'i}a-Perera, ``Overlap-aware
  diarization: Resegmentation using neural end-to-end overlapped speech
  detection,'' \emph{IEEE International Conference on Acoustics, Speech and
  Signal Processing (ICASSP)}, pp. 7114--7118, 2020.

\end{thebibliography}

\end{document}